\documentclass[twocolumn,  preprintnumbers,amsmath,amssymb]{revtex4}
\usepackage{graphicx}
\usepackage{dcolumn}
\usepackage{bm}
\begin{document}

\title{Interactions between colloidal particles induced by polymer brushes grafted onto the substrate}

\author{Kang Chen and Yu-qiang Ma}
\altaffiliation[ ]{Author to whom correspondence should be
addressed. Electronic mail: myqiang@nju.edu.cn.}

\address{National Laboratory of Solid State Microstructures,\\
Nanjing University, Nanjing 210093, China}

\begin{abstract}
We investigate the interaction energy between two  colloidal
particles on or immersed in nonadsorbing polymer brushes grafted
onto the substrate as a function of the separation of the
particles by use of self-consistent field theory calculation.
Depending on the colloidal size and the penetration depth, we
demonstrate an existence of repulsive energy barrier of several
$k_{B}T$,  which can be interpreted by separating the interaction
energy into three parts: colloids-polymer interfacial energy,
entropic contribution due to ``depletion zone" overlap of
colloidal particles, and entropically elastic  energy of grafted
chains by compression of particles. The existence of repulsive
barrier which is  of entirely entropic origin, can lead to kinetic
 stabilization of the mixture rather than depletion flocculation or phase separation.
 Therefore, the present result may suggest an approach to control
the self-assembling behavior of colloids for the formation of
target structures, by tuning the colloidal interaction on the
grafting substrate under appropriate selection of colloidal size,
effective gravity (influencing the penetration depth), and brush
coverage density.

\end{abstract}
\maketitle

Assembling colloidal nanoparticles into ordered structures is
essential to the fabrication of nano-materials with high
performance in optical, electrical, and mechanical properties.
\cite{1,1qq,2,888,8885,7,77,777,7777,8,8qq,8qqq,linn,howen} Due to
the small length scale and interaction energy, the ordered
structures normally are realized by self-assembly of nanoparticles
in solution or at an interface for the required motion to reach
minimum energy state.\cite{2} The mechanism of such
self-assembling behavior greatly relies on the effective
interaction between particles mediated by the environment such as
solvent, substrate, and external field. The fundamental element is
obviously the effective interaction between two individual
colloidal nanoparticles as a function of their distance, and the
results will be  helpful to our understanding of the collective
assembling behavior of large amount of particles. Usually, the
interaction between large colloidal particles in a solution of
small ones or nonadsorbing polymers is strongly influenced by the
entropic excluded-volume(depletion) forces which were first
recognized theoretically by Askura and Oosawa(AO).\cite{999} Such
an effect can be successfully used to explain  phase separation in
colloid-colloid
 and colloid-polymer mixtures. \cite{d} Interestingly, Lin et al.\cite{linn} combined
 entropic depletion and patterned surfaces to
  study the self-assembly of colloidal spheres on periodically patterned
  templates. Many theoretical and experimental works have already been performed to examine
the entropic depletion interactions between colloidal particles
 in  different suspensions  as depletion
 agents.  The evolved systems include  colloid/polymer mixture, \cite{t17,t1755,t19,t20} colloids
 in DNA solutions \cite{t21} or nematic liquid crystal/rigid rod background, \cite{t22,t23,t15,t18,t28,t29,t30}
 confined colloidal systems \cite{t24,t25} and binary hard-sphere mixtures. \cite{t16,t26,t27}
In colloid/polymer suspensions, when the distance between sphere
surfaces or between sphere surface and the wall is smaller than
the diameter of polymer coils, there will be an attractive
potential which depends strongly on the polymer concentration. The
topology of phase diagram depends on the ratio of polymer radius
of gyration to the radius of colloidal particles. Usually, in the
colloid limit with small polymer to colloid size ratios, depletion
interactions can be well described in pairwise form. On the
contrary, the protein limit for larger polymer sizes requires the
incorporation of many-body contributions.\cite{t20} Direct
measurement of depletion of colloids in semidilute DNA solutions
has revealed the range of ``depletion zones" proportional to the
correlation length.\cite{t21}  Recently, Fredrickson et al.
\cite{t31} investigated the potential of mean force between two
nanoparticles in a symmetric diblock copolymer matrix using
self-consistent-field (SCF) theory.

  On the other hand, the grafting of polymer chains to  surfaces
is a widely used method to modify their properties including
adhesion, lubrication, and wetting behavior, and has many useful
applications such as colloidal stabilization, polymeric
surfactant, biocompatibility, and drug carriers. SCF theory
formulated in bispherical coordinates was developed to study the
interaction between surface-grafted spherical assemblies, and the
results show hints to the conditions for colloidal stabilization.
\cite{kawa,kawa1} Recently, polymer-coated surface has become an
alternative substrate for the assembly of colloidal nanoparticles
and metal nanocrystals because of the``softness" property of
interfaces and the nanoscale characteristic length. The entropic
elastic energy of such polymer brushes will play an important role
in colloidal assembly since the typical energies of both
polymer-chain stretching and colloidal assembly are comparable to
thermal energy. For example, the grafted polymer always exerts a
repulsive entropic force on incoming particles, and most works
focused on the interaction between polymer brushes and individual
incoming particles and how to prevent the adsorption of colloids
such as proteins onto surfaces under various grafting density,
chain length, and interactions between chains and surface.
\cite{t14,9,9ee,9eee,9eeee,9eeeee,9ed} Some experiments have
adopted polymer-grafted surface to control the assembly of
nanoparticles. \cite{7,77,777,7777} Assemblies of nanoparticles
with number density gradients in two and three dimensions are
achieved. However, to the best of our knowledge, there have been
no systematic theoretical and experimental studies into the
interaction energy between two colloidal particles  on or immersed
in the polymer brushes, which is vitally important to
understanding the assembling behavior of many colloidal particles
on the polymer-grafted substrate.

In the present paper, we propose a theoretical model to study the
effective interaction between two hard colloidal particles  on or
immersed in polymer brushes with different penetrating depths and
particle's radii  comparable to the natural size of polymer coils.
We try to present a detailed and complete picture on such unique
interaction potential, in order to understand the kinetic
 stabilization(i.e., colloidal dispersions) and    flocculation of colloids
 due to the presence of grafting polymer chains. The results are expected to be helpful to assembling experiments of
 particles on such a substrate with end-grafted polymers.

\begin{figure}
\includegraphics[width=7.cm]{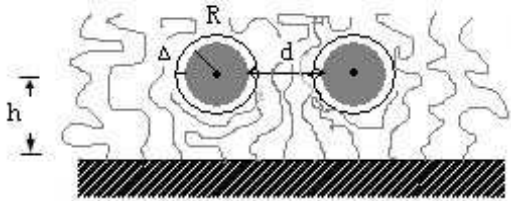}
\caption{ A schematic   of the two colloidal particles immersed in
polymer brushes. }
\end{figure}

In our two-dimensional model, the system is placed in xz plane and
the substrate is horizontally fixed at $z=0$ with $n_{\alpha}$
uniformly grafted polymer chains (see Fig.1). On the top of the
brush is the  polymer solvent composed of free  chains. Two
colloidal particles are modelled by hard circles of radius R
located a distance $d+2R$ apart and a distance
 h from the substrate; \cite{xmz} there are no
 attractive or repulsive interactions between them. Each colloidal particle
 is surrounded by a depletion zone of thickness $\Delta$, where the concentration of polymers
  becomes depleted compared to that of the bulk solution.\cite{d,t21}
All polymer chains and colloids are allowed in the region $0\leq
z\leq Z_{max}$. Here, $Z_{max}$ is large enough to avoid the
influence of brushes and particles, and as $Z$ approaches
$Z_{max}$, the free ``solvent" chains   attain their bulk
properties.\cite{12} The volume of the system $V$ is $L_x\times
Z_{max}$, where $L_x$ is the lateral length of the surface along
the $x$ axes. The grafting density is defined as $\sigma
=n_{\alpha}/L_x$. All polymer chains are of the same
polymerization index $N$ and flexible with the same statistical
length $a$, and incompressible with a segment volume $\rho
_0^{-1}$.
 The probability distribution for  molecular
conformations of a Gaussian chain $\alpha$  is assumed to take the
Wiener form $
\textit{P}[\mathbf{r}_{\alpha}(s)]\varpropto\exp[-\frac{3}{2a^{2}}\int_{0}^{1}ds|\frac{d\mathbf{r}_{\alpha}(s)}{ds}|^{2}]
$, where ${\bf r}_\alpha(s)$ denotes the position of segment s on
chain $\alpha$.
 In every calculation, the positions of two colloids are fixed,
and the local volume fraction $\varphi_p(\bf r)$  of   the
colloidal particle of radius $R$ with center at position $\bf r_c$
is described by taking into account the depletion
interaction\cite{13,ham} between colloids and polymer chains,
\[
\varphi_p(\bf r)=\left\{\begin{array}{cc} 1, & |{\bf r}-{\bf
r}_c| \leq R, \\
\frac{1+cos((|{\bf r}-{\bf r}_c|-R)\pi/\Delta)}{2}, & R \leq |{\bf
r}-{\bf r}_c| \leq R+\Delta, \\
0, & R+\Delta \leq |{\bf r}-{\bf r}_c|. \end{array} \right .
\]
We take the incompressibility constraint that
$\varphi_{\alpha}({\bf r})+\varphi_{\beta}({\bf r})+\varphi_p({\bf
r})=1.0$, which will ensure the exclusion of polymer chains from
the depletion zones surrounding the two colloidal particles. Here,
$\varphi_{\alpha}(\bf r)$ and $\varphi_{\beta}(\bf r)$ represent
local volume fractions of grafted and free chains, respectively.
The size $\Delta$ of depletion zone is determined by depletion
agents in the surrounding environment, and is independent of the
radius $R$ of colloids. For highly concentrated polymer solutions
the ``depletion zone" thickness is chosen to be about the size of
a free segment\cite{t21}, and here we set $\Delta=a$.\cite{abcz}
We define Flory-Huggins interaction parameters between
brush-solvent($\chi_{\alpha \beta}$), brush-particle
($\chi_{\alpha p}$), and solvent-particle ($\chi_{\beta p}$), and
take
 $\chi_{\alpha \beta}N=0.0$ and $\chi_{\alpha p}N=\chi_{\beta p}N=20.0$  since we assume
that all the polymers have the same chemical nature, while the
colloidal particles are insoluble to polymers.  We use  the grand
canonical form of  SCF theory, which has been proven to be
powerful in calculating equilibrium morphologies in polymeric
system,\cite{12,13,1318,14,14a,14aa,14aaa,14aaaa,14aaaaa,14aaaaaa,15,1515}
to deal with polymer solvents and brushes.

 The grand canonical partition function \cite{12} for the system
can be written as
\begin{equation}
Z_\mu=\sum_{n_\beta =0}^{\infty} e^{n_\beta N \mu}%
Z_{n_\alpha , n_\beta}\;,
\end{equation}
 where $\mu$ is the chemical potential per
segment of solvent chains.  $Z_{n_\alpha,n_\beta}$ is the
canonical partition function for $n_\alpha$ grafted chains and
$n_\beta$ free polymers:

\begin{eqnarray}
Z_{n_\alpha, n_\beta} &=&\frac{1}{n_\alpha !} \frac{1}{n_\beta !}%
\int\prod_{\alpha =1}^{n_\alpha}%
\textit{D}\mathbf{r}_{\alpha}(s)\textit{P}[\mathbf{r}_\alpha(s)]
\prod_{\beta%
=1}^{n_\beta}\textit{D}\mathbf{r}_{\beta}(s)\nonumber\\
&&\textit{P}[\mathbf{r}_\beta(s)]\delta[1-\widehat{\varphi}_{\alpha}-\widehat{\varphi}_{\beta}
-\varphi_p]\exp[-\frac{\nu}{k_B T}]\nonumber\\
&&\prod_{\alpha=1}^{n_\alpha} \delta(\mathbf{r}_z^\alpha(0))\;\;,
\end{eqnarray}
where the first $\delta$ function enforces incompressibility and
the second ensures the anchoring of the chain ends on the
substrate. $k_B$ is the Boltzmann constant, $T$ is the
temperature, and $\nu$ is the interaction energy.
$\widehat{\varphi}_{\alpha}$ and $\widehat{\varphi}_{\beta}$ are
corresponding operators to local concentration $\varphi_{\alpha}$
and $\varphi_{\beta}$. The end of brush chains can move on the
substrates, although the total number of chains on surfaces is
fixed(we take $\sigma =0.25$). This is so-called liquid brushes,
contrary to solid brushes where  the immobile chains are anchored
onto surfaces\cite{9eee}. Here we consider the liquid brush case
because of its wide-ranging applications
 in colloidal self-assembly and biological
organization including cell adhesion and interactions between
polymer-coated membranes and  proteins(or cells)\cite{9eee}.
  The SCF theory gives the free
energy
\begin{eqnarray}
\frac{N \textbf{F}}{\rho_0 k_B T V} &=&-\phi\ln(\frac{\mathbf{Q_{\alpha}}}{\phi V})%
-\frac{N}{\rho_0 V} e^{N \mu} \mathbf{Q_{\beta}}
+\frac{N \nu}{\rho_0 k_B TV} \nonumber\\
&&-1/V \int \textit{d}\mathbf{r}
[\xi(1-\varphi_{\alpha}-\varphi_{\beta}
-\varphi_p)\nonumber\\
&&+w_{\alpha} \varphi_{\alpha}+w_{\beta} \varphi_{\beta}]\;\;,
\end{eqnarray}
where $\phi$ is the overall volume fraction of brushes.
$\mathbf{Q_{\alpha}}=\int \textit{d}\mathbf{r}
q_{\alpha}(\mathbf{r},s)q_{\alpha}^{\dag}(\mathbf{r},s)$
represents the single chain partition function of grafted chains
subject to the field $w_{\alpha}$, and $\mathbf{Q_{\beta}}=\int
\textit{d}\mathbf{r}
q_{\beta}(\mathbf{r},s)q_{\beta}(\mathbf{r},1-s)$ is the partition
function for solvent under the field $w_{\beta}$. The end-segment
distribution functions $q_{i}(\mathbf{r},s)$
  and $q_{i}^{\dag}(\mathbf{r},s)$
(or $q_i(\mathbf{r},1-s)$) stand for the probability of finding
the $s^{th}$ segment at position $\mathbf{r}$ respectively from
two ends of grafted (or free) chains. The $q_i$ satisfies a
modified diffusion equation $\frac{\partial q_i}{\partial s}=
\frac{N a^2}{6} \nabla^2 q_{i}-w_{i}(\mathbf{r})q_i$. $q_{i}^\dag$
meets the same diffusion equation but with the right-hand side
multiplied by -1. The interaction $\nu$ is given by
\begin{equation}
\frac{N   \nu}{\rho_0 k_B T V}=\frac{1}{V} \int \textit{d} \mathbf{r} [\chi_{\alpha \beta} N \varphi_{\alpha}
\varphi_{\beta}%
+\chi_{\alpha p} N \varphi_{\alpha} \varphi_p +\chi_{\beta p} N
\varphi_{\beta} \varphi_p]\;\;.
\end{equation}

\begin{figure}
\includegraphics[width=8.cm]{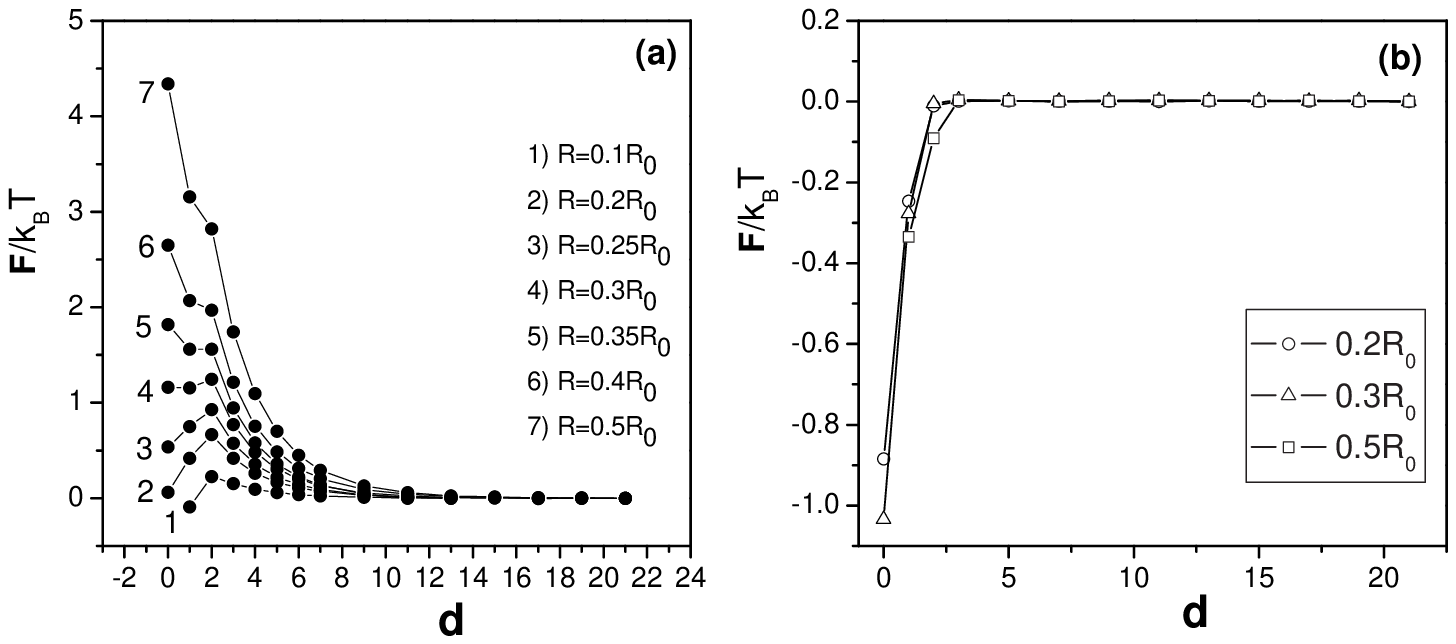}
\includegraphics[width=8.cm]{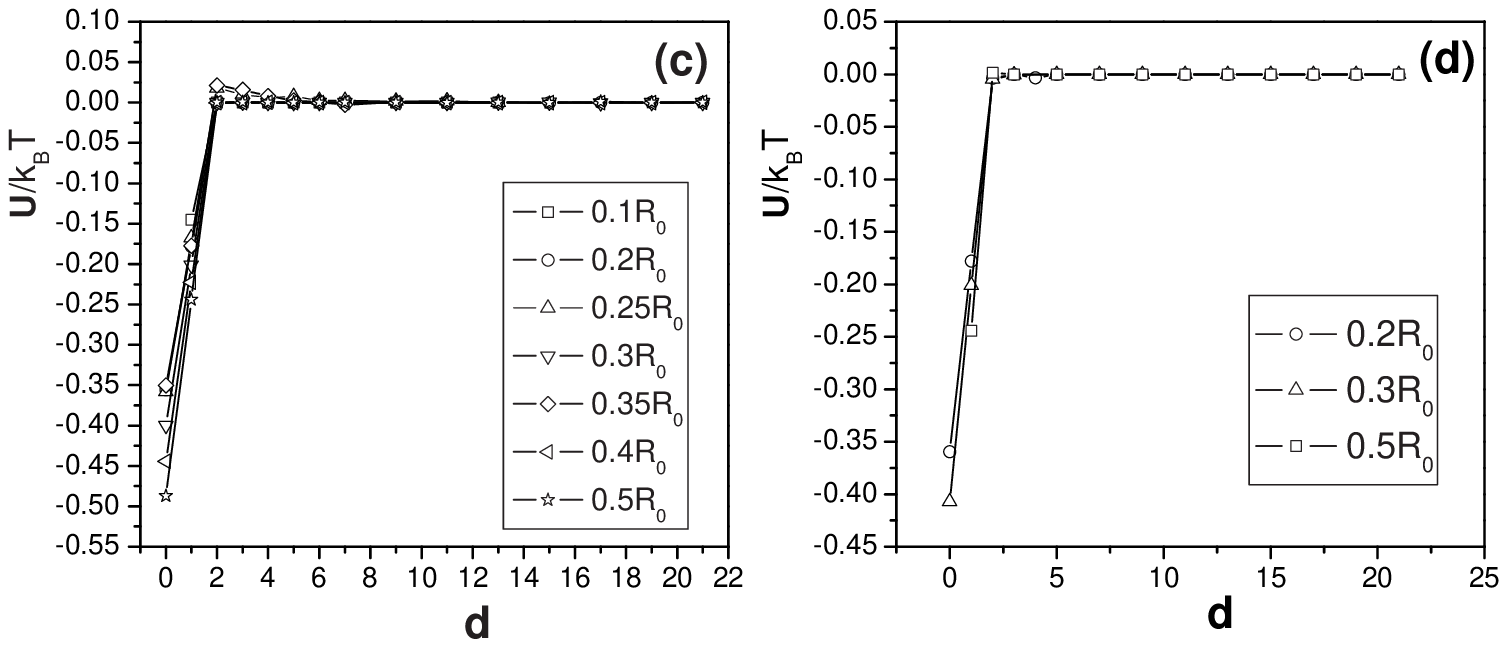}
\includegraphics[width=8.cm]{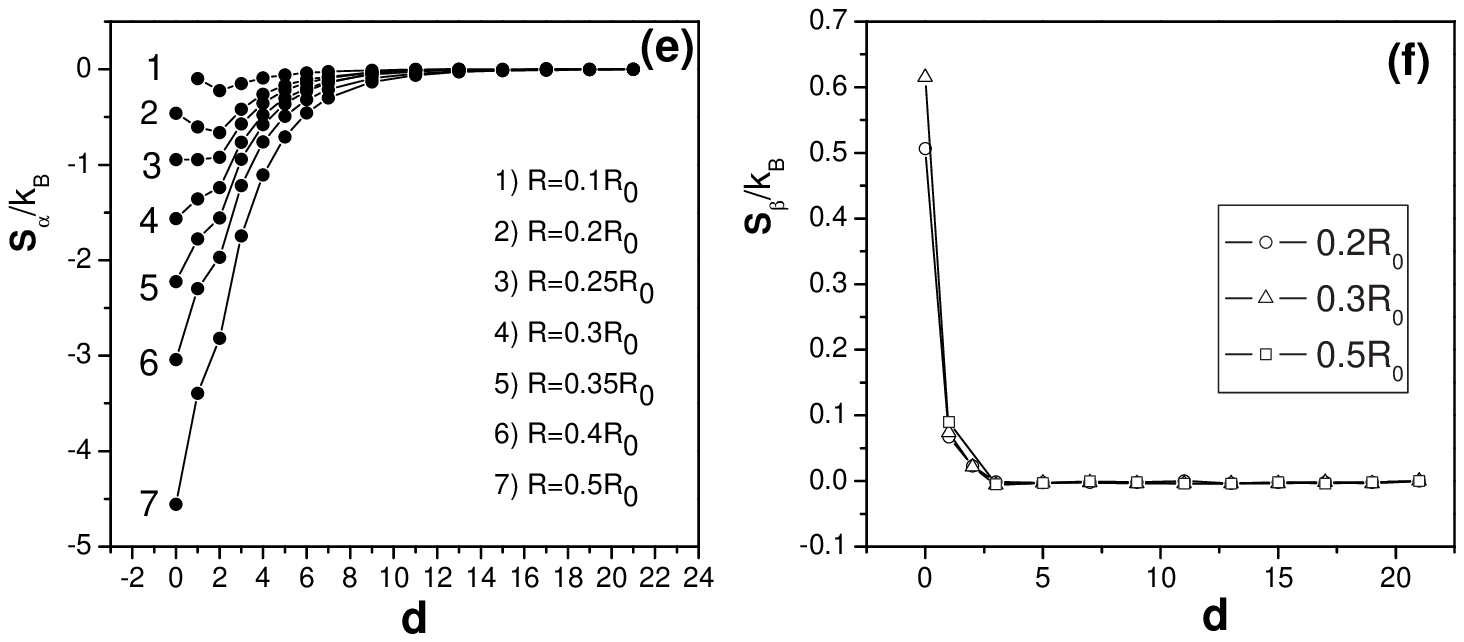}
\caption{Comparison of free energy   in the present brush
system(a) and in pure polymer melt(b), interfacial energy in  the
present brush system(c) and in pure polymer melt(d), and brush
entropy in the present system(e) and free chain entropy in the
pure polymer melt(f).}
\end{figure}
In SCF theory, the fields and densities are determined by locating
saddle points in the free energy expression Eq.(3). The resulting
equations can be solved self-consistently, and here we implement
the combinatorial screening algorithm of Drolet and
Fredrickson.\cite{15,1515} A periodical boundary condition for
$x$-direction is applied, while for the $z$-direction, the region
of $z<0$ is forbidden and the region of $z>Z_{max}$ is treated as
a bath of solvent. $Z_{max}$ is fixed to be $80a$, and the lateral
simulation box size $L_x=100a$ is large enough to avoid any
influence due to the presence of colloidal particles.

We first examine the interaction energy of two colloidal particles
(of radius R) deeply immersed in polymer brushes with particle's
radii smaller than or comparable to polymer coils. Figure 2(a)
gives the free energy  $F$ as a function of lateral separation $d$
between two colloidal particles, where the radii $R$ of particles
are chosen as $0.1R_0$, $0.2R_0$, $0.25R_0$, $0.3R_0$, $0.35R_0$,
$0.4R_0$, and $0.5R_0$ (from bottom to up), and the depth $h$ is
fixed at $12.0a$ which is smaller than the pure brush
height(around $25.0a$).  $R_0 \equiv a N^{\frac{1}{2}}$
characterizes the natural size of polymers, and we take $N=100$
here. We find that, with increasing colloidal particle's radius,
the free energy curves change from shapes of short-range
attraction and long-range repulsion into purely repulsive one. The
energy barrier height is compared to thermal energy  $k_{B} T$,
and the position of the peak shifts from   $d= 2.0 \Delta$ to $d=
0$. In contrast, the position of global minimum of free energy
shifts to larger values of d, indicating a transition of favorite
colloidal aggregation to colloidal dispersion. We also find that
there may exist a weak double-peak behavior at middle particle
sizes (around $0.3R_0$ $\thicksim$ $0.35R_0$), where  a local
minimum is located at $d = 1.0 \Delta$ and two peaks at $d = 0$
and $2.0 \Delta$. For larger particles, the free energy decreases
with $d$, but at $d = 2.0 \Delta$, an inflexion point still
exists. For comparison, we calculate the free energy curves of
corresponding particles in pure polymer melts instead of polymer
brush (see Fig.2(b)). On the contrary, we can see that, in polymer
melt, the pair potential between colloidal particles is
monotonically attractive in spite of different sizes and the
interaction is short-range (about $2.0 \Delta$) with distance.

\begin{figure*}
\includegraphics[width=15cm]{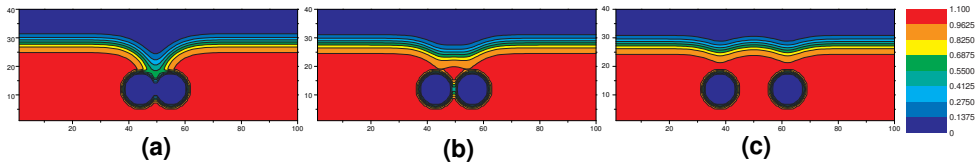}
\caption{  Concentration distributions of brush chains due to the
presence of two  colloidal particles with size $R=0.5R_0$ and
depth $h=12.0a$. (a) $d=0$, (b) $d=2.0 \Delta$, and (c) $d=13.0
\Delta$.}
\end{figure*}
In order to highlight the free energy feature shown in Fig. 2(a),
we examine the free energy $F$ of the colloidal particles in
polymer brushes which is composed of three main contributions:
interfacial energy between colloidal particles and polymer chains,
entropy contribution due to space overlap of depletion zones
(depletion potential), and entropically elastic stretching energy
of brush chains due to the compression of particles. The
interfacial energy is the same either for the particles immersed
in polymer brushes (Fig.2(c))
 or in free polymer melt (Fig.2(d)) due to the same chemical nature of
grafted and free chains. The interfacial energy curves   exhibit a
turning point at $d = 2.0 \Delta$ because the depletion zones
surrounding the two particles will overlap when the distance $d$
is smaller than $2.0 \Delta$, which will decrease the total
interfacial area between colloidal particles and polymer chains.
Figure 2(f) shows the entropy curves of free polymer chains when
the colloidal particles are immersed in polymer melt.
 The free chain entropy change with the two-particle separation
 $d$  mainly arises from space overlap effect of depletion zones.
  When $d \leq 2.0\Delta$, two colloidal particles approach
each other, and their depletion zones begin to overlap. This will
increase the total volume accessible to the free polymer chains,
leading to the increase of their entropy. Notice that in Fig.
2(f), slight decreases of the entropy appear for $d \geq 2.0
\Delta$, and the entropy curves reach their minimum
  at   $d = 3.0 \Delta$. Such a behavior is due to the decrease of conformational
entropy of polymer chains confined between the two-particle
surfaces, because when $d$ arrives at $3.0 \Delta$ the free
polymer chain begins to be permitted through the region between
two particle' surfaces. When the colloidal particles are immersed
in polymer brushes, the entropy of grafted chains will include
both the contributions of space depletion effect   and
entropically elastic effect of brushes (see Fig.2(e)). On the one
hand, the entropy due to chain compression reaches the minimum at
zero distance ($d=0$), and increases with the
 distance $d$. The reason is that when the particles
are far apart, the grafted chains will more easily  fill the upper
space through the interspace between the two-particle surfaces to
release the chain compression.   Figure 3 shows the brush
concentration distributions due to the presence of two colloidal
particles for three different distances $d=0$, $d=2.0 \Delta$, and
$d=13.0 \Delta$. The deformation of brushes is obviously weakened
with increasing the separation of colloidal particles.  On the
other hand, the space depletion effect leads to the   decrease of
chain entropy with $d$ for $d \leq 2.0 \Delta$ where depletion
zones surrounding the two particles overlap (see Fig.2(f)). It is
the competition between spatial depletion effect around hard
particles and entropically elastic energy of brushes, leading to a
possible appearance of the complex interaction for
 $0 \leq d \leq 2.0 \Delta$  and  the  shift of minimum values of entropy curves from
$2.0 \Delta$ to zero (Fig.2(e)). For small particles, the space
depletion effect will dominate over compression energy of brushes
and the curve has a minimum at $2.0 \Delta$. On the contrary, for
large particles the compression effect of brushes will become more
important and determines the minimum point at $d= 0$.  It is
actually interesting that   the brushes have large entropic
restoring force which easily controls the dispersions of colloids,
in contrast to non-adsorbing polymer solvent. Compared Fig.2(a)
with Fig.2(e), we find that the change of free energy   is almost
attributed to the entropic effect of brushes. On the other hand,
we see from Fig.2(c) that the interfacial energy mainly enhance
the formation of inflexion point at $d=2.0 \Delta$, and may lead
to the onset of double peaks of the free energy curves for middle
particle sizes.

\begin{figure}
\includegraphics[width=9cm]{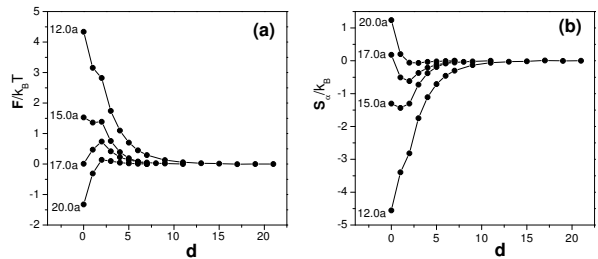}
\caption{Free energy(a) and brush entropy(b) curves of the system
 when two particles of size $R=0.5R_0$ are located in different penetration
depths.}
\end{figure}

To understand the brush entropy effects on effective interactions
between two colloidal particles, we now consider the influence of
colloidal penetrating depth $h$ on the free energy of the system
for fixed radius of  colloidal particles. Figure 4(a) and 4(b)
give the free energy and brush entropy curves, respectively, for
the particle size $R=0.5R_0$ with depths $h=
20.0a,17.0a,15.0a,12.0 a$. We can find that similar changes of
curve shapes for different $h$ values occur with the increase of
$d$. At high $h$, entropically compression energy of brush chains
is small compared with space depletion contributions. The free
energy has a minimum value at $d = 0$ where brush entropy reaches
its maximum. As $h$ is small, the compression contribution of
brushes becomes dominant, and the minimum value of free energy
tends to shift to large separation of the particles.
Correspondingly, the minimum point of brush entropy moves from
$2.0 \Delta$ to zero(Fig.4(b)). In particular, for middle depths
where the
 spatial depletion effect of hard particles in polymer solution may compete with
entropically elastic energy of brushes,  the richer interaction
potentials such as  repulsive barrier or double peaks may appear
for free energy curves.

\begin{figure}
\includegraphics[width=9cm]{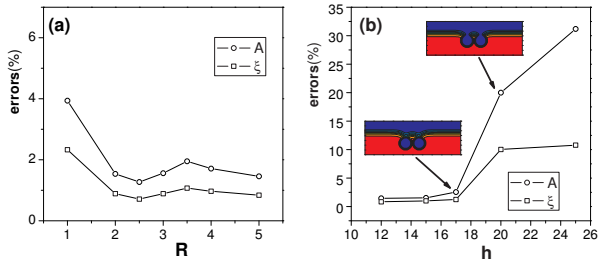}
\caption{ Exponential-decay function fitting errors for amplitude
$A$ and decay length $\xi$. (a) errors for two colloidal particles
of different sizes immersed in polymer brush; (b) errors for two
colloidal particles of size $R=0.5R_0$ under different penetrating
depths $h$.}
\end{figure}

To further examine the long-range force features due to
entropically elastic effect of  brushes, finally we define an
exponential-decay function of free energy as $C_0+A \; exp
(-\frac{d}{\xi})$ with increasing the two-particle separation d.
Figure 5 shows the fitting errors in percentage for amplitude $A$
and decay length $\xi$ as functions of colloidal particle size and
colloidal penetrating depth $h$, respectively. We find from
Fig.5(a) that when colloidal particles are immersed in brushes,
the errors are very small in spite of different particle sizes.
This means that the long-range particle-particle interaction
mediated by polymer brushes has an excellent exponential-decay
repulsive tail. Figure 5(b) provides the corresponding
exponential-decay fitting errors in percentage for the long-range
behavior of free energy curves as a function of penetration depth
$h$. We see that the errors, which reflect the degree of agreement
with the exponential decay behavior, have a sharp transition
between $h=17.0a$ and $20.0a$ where the particles start to reach
brush-solvent surface(see the inset of Fig.5(b)). The result
further addresses the fact that the deformation of brushes drives
the occurrence of  the exponential decay of long-range repulsions.

 In summary, we have revealed a complex interaction
between colloidal particles mediated by polymer brushes. Under
different colloidal sizes and penetrating depths, we  obtain
complex attractive and repulsive interaction behaviors as a
function of colloidal particles' separation, due to the complex
interplay among entropically stretching energy of grafted
polymers, entropic effects due to ``depletion zone" overlap of
hard particles, and the particles-polymer interfacial energy. The
short-range attraction is mainly attributed to depletion effects
of   particles in polymer solutions, whereas the long-range
interaction between penetrating particles is found to be a typical
result of brush entropy effect, which has a well-fitted
exponential-decay repulsive tail. By simply using the grafting
substrate, the present result indicates a possibility for
governing the tendency of colloidal particles to flocculate into
multi-particle aggregates or to remain dispersed, and may have
important implications for industrial   processes, especially for
biological systems.

Finally, we should stress that in the present case, dimensionality
is also an important factor. Here, the two-dimensional results can
be effectively used to describe the interaction between two
parallel long colloidal rods in three-dimensional experiment
systems. However, for the case of colloidal spheres in
three-dimensional systems, the deformation of grafted chains would
be weakened (or larger colloidal spheres are needed to yield the
equivalent deformation of grafted chains as in the two-dimensional
case). Then, the peaks of the free energy curves(for example, in
Fig.2(a)) are expected to decrease under the same parameters, and
the transition between attractive and repulsive interactions will
finally happen at larger colloidal radius. Additionally, we also
point out that the free energy difference caused by liquid and
solid brushes will greatly depend on the grafting density of
brushes. For low grafting densities,
 the  chains of liquid brushes
can easily relieve the deformation imposed by the colloidal
particles by moving out of the region between colloids, which
favors the effective attraction between colloidal particles at
short distances.  On the contrary, for solid brush case where the
end of chains  cannot move away and the grafting density keeps
uniform, the total free energy and brush entropy will depend on
the relative lateral positions between the colloidal particles and
the grafting end-points of chains on the substrate. For example,
the existence of immobile grafting chain between the colloidal
particles will enhance the entropy repulsive barrier to
particle-particle interaction at short  distances. This means that
the resulting curves of free energy and brush entropy obtained for
mobile and immobile grafted chains will be obviously different.
However, at high grafting density, even the mobile grafted chains
cannot move easily due to the incompressibility constraint, i.e.,
the grafting density almost keeps uniform on the substrate for
mobile grafted chains. On the other hand, the brush height
increases with increasing the grafting density, and thus the
grafted end effect on particles becomes less important. In this
case, there is no distinct difference between liquid and solid
brushes, and the resulting free energy curves will be similar.

This work was supported by  the National Natural Science
Foundation of China, No. 10021001, No. 10334020, and  No.
20490220.


\begin{references}
\bibitem{xmz} For simplicity, here we adopted a two dimensional model,
namely  the colloidal particles can be taken as two  circles.
However, the present model is easily extended to three-dimensional
case where the   colloidal particles can take spherical or other
shapes.
\bibitem{ham}In Ref. 46, Matsen use
a simple equation (Eq.(2)) to describe the depletion interaction,
however, the work  focused on the morphologies of copolymers
confined between two fixed plates(i.e., the effective interaction
between two plates due to the existence of copolymers between them
is not considered). In contrast, here the depletion zone is
introduced to emphasize the study of depletion interaction between
free colloids arising from surrounding polymer chains. For the
same reason, the details regarding the segment profile at the two
surfaces are not important, and so we will choose a similar form
by Matsen,  Ref. [46].
\bibitem{abcz} We also calculated larger $\Delta$ values of depletion zone, and found that the physical features
are qualitatively the same except the small shifts of the peak of
free energy curves.
\end{references}
\end{document}